# The "weighted ensemble" path sampling method is statistically exact for a broad class of stochastic processes and binning procedures

Bin W. Zhang,[1] David Jasnow,[2] and Daniel M. Zuckerman [*,1]

[1]*Department of Computational Biology, School of Medicine,*
*University of Pittsburgh, Pennsylvania 15260*

[2]*Department of Physics & Astronomy,*
*University of Pittsburgh, Pittsburgh, Pennsylvania 15260*

## Abstract

The "weighted ensemble" method, introduced by Huber and Kim, [G. A. Huber and S. Kim, Biophys. J. 70, 97 (1996)], is one of a handful of rigorous approaches to path sampling of rare events. Expanding earlier discussions, we show that the technique is statistically exact for a wide class of Markovian and non-Markovian dynamics. The derivation is based on standard path-integral (path probability) ideas, but recasts the weighted-ensemble approach as simple "resampling" in path space. Similar reasoning indicates that arbitrary nonstatic binning procedures, which merely guide the resampling process, are also valid. Numerical examples confirm the claims, including the use of bins which can adaptively find the target state in a simple model.

[*] Electronic mail: ddmmzz@pitt.edu

## I. INTRODUCTION

In 1996, Huber and Kim introduced a path sampling method which they dubbed "weighted ensemble Brownian dynamics" (WEBD) simulation. [1] Their focus was the diffusively controlled binding of a ligand to a receptor, and the WEBD method was designed to guarantee (statistically) that some ligands would not simply wander away while also permitting the correct calculation of kinetic rates. The method was later extended to configuration space in the folding of coarse-grained proteins. [2] Our own use of the approach studied a conformational transition, and we demonstrated that the method produces not only kinetic information but also the full transition path ensemble. [3] All these studies considered only Markovian dynamics. We note that "forward flux" sampling [4,5] uses a strategy similar to WEBD; there are also related steady-state methods, [6,7] and Monte Carlo approaches. [8] Furthermore, there are other rigorous path sampling and generating techniques. [9–19]

The aim of the present note is to establish or expand upon two important facts about the weighted ensemble (WE) method. (i) *The WE method produces unbiased results for a broad class of stochastic dynamics, including non-Markovian processes.* Thus the "BD" in Huber and Kim's original title was overly restrictive. Non-Markovian dynamics can arise, for instance, in self-avoiding walks [20] or as a means for accounting for the effects of degrees of freedom which have been suppressed — for instance, the hydrodynamic effects of solvent molecules not explicitly modelled. [21] (ii) Further, the WE method can use arbitrary bins which can change over time. This point was made in the original WE paper [1] and re-iterated in our own work. [3]

To demonstrate (i) and (ii), we will show that WE is simply a resampling process - i.e., it generates an alternative but equivalent statistical sample of trajectories at occasional time points. Further, this leads to no bias in the future evolution of the system, subsequent to resampling. Because all dynamical information, including the configurational distribution, derives from the trajectory ensemble, WE produces fully unbiased results. In WE simulation, the resampling is controlled by the choice of bins. Because of the flexibility intrinsic to resampling, however, there is great flexibility in bin selection - including the freedom to dynamically re-define bins on the fly. As shown below, it is even possible to re-define bins adaptively — without using knowledge of the target state — which eventu-ally lead to successful transitions. This opens up the possibility to explore configurational changes not already described by experimental structures.

## II. THEORY

### A. The "weighted ensemble" procedure

The weighted ensemble procedure of Huber and Kim [1] is straightforward to describe. First, the space of interest (real or configurational) is divided into regions called "bins." The choice of bins is arbitrary and may change during a simulation, as discussed below. For concreteness, however, it may be presumed that the bins are static and chosen so that some or all bins lead sequentially to a "target state" of interest, which might be a binding site in real space or a configuration-space state. Let $N$ be the number of bins.

In conventional WE simulations $M$ stochastic trajectories are initiated from one bin and each is assigned a weight (probability) of $1/M$. Any distribution of initial configurations may be used, whether confined to a single bin or not. The trajectories are run for a short interval of time and stopped, at which point the current bin of each trajectory is determined. Trajectories arriving to new bins are split into identical daughter trajectories sharing the weight and history of the parent trajectory. Typically, $M$ trajectories are created in each newly visited bin, and each daughter inherits a suitable fraction of the parent's weight (e.g., $1/M$ if only one trajectory arrives). All trajectories are then restarted and run for another short interval of time, after which the process of splitting is repeated as necessary. Whenever more than $M$ trajectories occupy the same bin, some are "killed" by a probabilistic resampling process (see below), maintaining a total weight of one and no more than $NM$ trajectories based on $N$ bins. The whole process is repeated until the desired information is obtained.

The WE procedure generates a weighted trajectory ensemble, which is rich in information. For instance, the transition rate from the initial state to the target can be calculated based on the probability arriving in the target state as a function of time; in simple cases, the arrival probability per unit time rapidly plateaus to the value of the rate.[1,3] (More generally, the rate can also be calculated based on a steady-state WE procedure, as we have shown elsewhere. [22]) The trajectory ensemble also includes information on more detailed transition probabilities and the evolving configurational distribution, both of which can be used to analyze or identify "structural" features [3] such as intermediate states.

### B. Resampling

The first goal is to get a feel for "resampling", [23,24] which is a method for generating an alternative sample of a probability distribution, given some initial sample. Consider a simple static example, where we initially generate, say, 100 numbers distributed according to a Gaussian. We could now resample this distribution in a number of ways which preserve the distribution. For instance, we could (a) discard 50 numbers at random or (b) duplicate each number exactly twice. It is clear that such processes are statistically correct by a "repeated simulations" argument: if we average over many repetitions of the sample-generation and resampling process — e.g., repeated generation of many sets of 50 numbers by process (a) — then we will generate a Gaussian distribution to any desired precision.

Resampling can also be done in a non-uniform way, by keeping track of weights. Consider the same set of 100 "Gaussian numbers", with zero mean. In the initial "ordinary" statistical sample, we can say that each number has a constant weight which can be set to one without loss of generality. We could randomly discard half the numbers on the left ($x < 0$) and assign each of the remaining left-side numbers a weight of two. Again by the repeated simulation argument, this is correct resampling. Alternatively, we could duplicate all the numbers on the right ($x > 0$), and assign each a weight of one-half. More strangely at first glance, we could simultaneously do both resampling procedures, which would lead to a sparse sample on the left and a dense sample on the right. The repeated simulation argument also demonstrates the validity of this procedure.

*Resampling can be defined as a process which creates an alternative, but statistically equivalent, sample of a fixed probability distribution based on an initial sample.* The resampled sample will contain only a subset of elements in the original sample but with potentially different frequencies and compensating weights. The new sample can be either larger or smaller than the original, as desired.

Importantly, the elements in a sample — which can be resampled — can have arbitrary dimensionality. They can be vectors or "objects" of any kind and, in particular, a vector can represent a *dynamical* process/history. For instance, a vector can represent the sequence of configurations in a discretized trajectory. Distributions of trajectories will now be considered more carefully.

### C. Statistical description of discretized, stochastic trajectories

We would like to construct a probabilistic description of stochastic processes, and in particular, the probability distribution of trajectories.[9,12,13] Because our goal is to perform more effective computer simulations, we will restrict ourselves to discretized dynamics. That is, for our purposes, the exact dynamics will be enacted by a computer simulation of the form

$$\mathbf{x}_j = f(\mathbf{x}_0, \mathbf{x}_1, ..., \mathbf{x}_{j-1}) \tag{1}$$

which indicates that the present system configuration $\mathbf{x}_j$ is a (probabilistic) function of the full history of the trajectory. In other words, in the notation of [25], $\mathbf{x}_j$ is chosen from the conditional probability distribution $p_{1|k}(\mathbf{x}_j|\mathbf{x}_{j-k}, \mathbf{x}_{j-k+1}, ..., \mathbf{x}_{j-1})$, which depends on the previous $k$ steps of the trajectory. We are restricting ourselves here to processes which are homogeneous in time, i.e., the function $f$ (or the conditional probability) is independent of the time step $j$.

The important special case of a Markov process corresponds to $k = 1$. For $k > 1$, an operational rule is required to initialize the trajectory: if $j < k$, the function $f$ must embody some rule for using the distribution $p_{1|k}$, e.g., setting all earlier $\mathbf{x}_i$ (for $i < 0$) to some arbitrary value(s).

Such a dynamical processes implies that we can construct the full probability distribution $P^{\text{path}}$ for $n$-step trajectories as the product of the initial distribution $p_1(\mathbf{x}_0)$ and subsequent conditional probabilities $p_{1|k}(\mathbf{x}_j|\mathbf{x}_{j-k}, ..., \mathbf{x}_{j-1})$ That is, we have

$$P^{\text{path}}(\mathbf{x}_0, ..., \mathbf{x}_n) = p_1(\mathbf{x}_0) \prod_{j=1}^{n} p_{1|k}(\mathbf{x}_j|\mathbf{x}_{j-k}, ..., \mathbf{x}_{j-1}). \tag{2}$$

This formal - but explicit - equation for the probability distribution of trajectories shows that for the broad class of stochastic dynamics governed by Eq.(1), the distribution of trajectories can be considered a high dimensional "equilibrium" distribution.[9,12,13] Indeed, it is this fact which permits the use of path integrals.

It is also of interest to consider the configuration-space distribution. This evolving distribution can be derived from Eq.(2) at any point in time simply by integrating over all possible histories. [26] That is, the distribution at time point $n$ which depends on the initial distribution, is given by (again in the notation of [25])

$$p_1(\mathbf{x}_n) = \int d\mathbf{x}_0...d\mathbf{x}_{n-1} P^{\text{path}}(\mathbf{x}_0,...\mathbf{x}_{n-1}) . \tag{3}$$

In the case of a continuous-time Markov process, this is the distribution that would be calculated from solutions of the Fokker-Planck equation. Note that the distribution $p_1$ in Eq.(3) is *not* the same as that in Eq.(2); rather we are following van Kampen's convention, [25] where the subscript refers to the number of variables in the argument and the conditionality if present.

### D. Resampling a distribution of trajectories

Because the distribution of trajectories $P^{\text{path}}(\mathbf{x}_0,...,\mathbf{x}_n)$ is an ordinary, albeit high-dimensional, distribution, it can be resampled just as described for "equilibrium" distributions. That is, given one sample of trajectories, we can generate another (statistically) correct sample using deletions and duplications of entire trajectories, provided we attach the correct weights to the remaining trajectories. The means for doing this are identical to that for the equilibrium case, and the specific procedure used in WE simulation is described later.

The distribution of *trajectories* at future times — subsequent to a resampling process — will also be correct. That is, if we resample the trajectory distribution at time step $n$ — preserving $P^{\text{path}}(\mathbf{x}_0,...\mathbf{x}_n)$ — the trajectories which evolve from the new sample via the dynamics (1) will also have the correct distribution at a later time step $n + m$. This can be seen formally by decomposing the full distribution according to the relation

$$P^{\text{path}}(\mathbf{x}_0,...,\mathbf{x}_n,\mathbf{x}_{n+1},...,\mathbf{x}_{n+m}) = P^{\text{path}}(\mathbf{x}_0,...,\mathbf{x}_n) \prod_{j=n+1}^{n+m} p_{1|k}(\mathbf{x}_j|\mathbf{x}_{j-k},...,\mathbf{x}_{j-1}) , \tag{4}$$

which is equivalent to propagating the dynamics via (1) from step $n$ to step $n + m$.

### E. WE as resampling

Weighted ensemble simulation simply performs occasional resamplings of the trajectory distribution. In operational terms, it follows a set of simultaneous trajectories simulated in parallel. Thus, the only difference between WE and brute-force simulation of this set of trajectories is that the WE method entails resampling at occasional time points. Between resamplings, the WE simulations employ ordinary brute-force dynamics. [1]

Resampling in WE is performed only among trajectories at the same point in time, and this is done in two ways. First, an entire trajectory may be replicated $M$ times, with each replica assigned a weight of $w_i/M$ if the original trajectory had weight $w_i$. Such replicas are assigned a history identical to that of the original trajectory. Second, two trajectories, $i$ and $j$ may be "combined": The full weight $w_i + w_j$ is assigned to one of the trajectories, with probability governed by the relative weights. The surviving reweighted trajectory retains its original history. Such combination is, in fact, simply a removal process, like those described in the examples above. The two types of resampling processes in WE, by construction, preserve the correct probability distribution for *trajectories* for the time at which resampling is performed. Therefore, as discussed above, the correct trajectory distribution is preserved at all future times.

Because the WE resampling preserves the distribution of trajectories, it also preserves the correct configurational distribution, which is derived from the trajectory distribution via Eq.(3).

### F. Resampling in WE simulation can be achieved with arbitrary, dynamically changing bins

In its simplest form, weighted ensemble simulation divides configuration space into a fixed set of bins or regions. Resampling is performed to ensure that the number of trajectories in each bin is equal (once a bin has been visited). For details, see, for example, Ref.[3].

The arguments above have emphasized that WE simulation is more general than originally thought. Specifically, we have argued that (i) resampling can be performed correctly in myriad ways, and (ii) that any correct resampling procedure will preserve the correct stochastic dynamics in a WE simulation. Thus, WE simulation can exploit alternative resamplings.

Most importantly, as noted by Huber and Kim in their original paper,[1] the bin choices which govern resampling in a WE study can be adjusted during the simulation. Arbitrary changes to the binning during a WE simulation serve only to generate different — but statistically correct — resamplings.

We now turn to numerical illustrations.

## III. NUMERICAL RESULTS

To confirm our previous conclusions by simulations, we consider several examples.

### A. Colored noise

We first study one-dimensional stochastic dynamics [27] governed by the overdamped Langevin equation,

$$\frac{dx}{dt} = \frac{F(x)}{\gamma} + R(t), \tag{5}$$

where $F(x)$ is the physical, conservative force and $\gamma$ is the friction constant. The noise $R(t)$ can be taken as Gaussian white noise with zero mean and correlation

$$\langle R(t)R(t')\rangle = \left(\frac{2k_B T}{\gamma}\right)\delta(t-t') = 2D\delta(t-t'), \tag{6}$$

or colored noise with zero mean and exponential correlation

$$\{\langle R(t)R(t')\rangle\} = D\lambda \exp(-\lambda|t-t'|), \tag{7}$$

where $\{\cdots\}$ denotes the average over the distribution of the initial value of $R(t)$ according to

$$P(R(0)) = \frac{1}{(2\pi D\lambda)^{1/2}} \exp\left[-\frac{R(0)^2}{2D\lambda}\right]. \tag{8}$$

Here $k_B$ is Boltzmann's constant, $T$ is the temperature, and $D$ is the diffusion constant. The inverse of $\lambda$ in Eq.(7) is the correlation time for the colored noise. If we use the simulation time step size $dt$ as unit, this timescale can be expressed as

$$\lambda^{-1} = s \times dt.  \quad (9)$$

In the case of colored noise (7), the process (5) is non-Markovian for the single coordinate $x$.[25] The particular choice of noise correlations (7), related to an Ornstein-Uhlenbeck process, can be generated using an auxiliary Markovian variable as a matter of convenience. However we consider only the single non-Markovian variable, $x$, in which case (5) represents a simple stochastic differential equation with additive noise.

Both WE method and brute force simulation are applied to study the duration of transition events[28,29] for double-well potential

$$U(x) = 5.0 k_B T [1 - x^2]^2, \quad (10)$$

with $F(x) = -\frac{dU(x)}{dx}$. The duration of a transition is defined to be the time interval between the last time the trajectory leaves $x = -1$ and the first time it reaches $x = 1$.[29] In the left graph of Fig. 1 we compare brute force and WE simulation results for different colored noises with $s = 1$, $s = 20$, $s = 50$ and $s = 100$. The two types of numerical results match very well. In the right graph of Fig. 1, we show the WE simulation results for colored noise with $s = 1$ and white noise, which are almost identical, as one would expect. Thus for any $s \gg 1$, the white and colored noise results are dramatically different, and WE method reproduces the effects in detail.

### B. Myopic self-avoiding walk

We consider a self-avoiding random walk on a two-dimensional "simple cubic" grid as a more extreme non-Markovian example: the "dynamics" (i.e., transition probabilities) at any step depend on the complete previous history. In particular, we study a "myopic self-avoiding walk", following the definition in Madras and Slade's book.[20] Walkers will always look one step forward before they move. If there are nearest neighbors of the current position which have never been visited, the next step will be chosen uniformly from them. Otherwise the next step will be chosen uniformly from those neighbors which have been visited least often. This definition ensures that the walk will not be trapped, and thus contains more dynamical flavor than the strictly non-intersecting, self-avoiding random walk. [20] As shown in the left panel of Fig. 2, two absorbing walls are placed at $x = -1$ and $x = 15$, and all the walkers start from the origin $(0, 0)$. Those which reach the right absorbing wall before being absorbed by the left absorbing wall are defined as successful transition events.

Brute force simulation and the WE method are applied to obtain the distribution of the durations of the transition events. The results are shown in the right panel of Fig. 2, and they match very well. In WE simulation $(10.865 \pm 0.015)\%$ of all the walkers reached the right absorbing wall successfully, which is in agreement with the brute force result, $(10.854 \pm 0.004)\%$.

It is noteworthy that when applied to self-avoiding walk simulations, the WE method is operationally very similar to the early work of Erpenbeck and Wall, [30] as well as to later refinements (e.g., Refs. [31,32]).

### C. WE method with random number of bins

The WE method can use arbitrary bins which can change over time. To check this important fact, the WE program was changed to randomly choose the number of bins $N$ from a range before every resampling (splitting and combination). Because such random binning is <u>not</u> selected to aid the efficiency of WE simulation, it should provide a good test for the <u>correctness</u> of WE under somewhat "adverse" conditions. The modified WE program is applied to study the duration of transition events for a Brownian particle in the double well potential described by Eq.(10). In the WE simulations, the region $x < -1$ is treated as the first bin and the initial state, the region $x > 1$ is the last bin and the final state. The area between them $-1 \leq x \leq 1$ is divided into $(N - 2)$ bins evenly. As shown in Fig. 3 the modified WE program with fluctuating bins yields excellent agreement with the result from static bins.

### D. WE method with adaptive Voronoi bins

The WE approach can adopt clustering ideas to divide multiple simulations into groups, and change the bins during the simulation for better performance; see also Refs.

[7,33,34]. In a particularly intriguing approach, the bins can be constructed and adjusted without using information about the target state. Such an adaptive strategy could be important in biomolecular simulations, where only a single state (e.g., experimental protein structure) is known, but the presence of other states is suspected.

The idea in adaptive WE simulation is for the bins to follow the evolving probability distribution. For example, the simulation can employ bins arising from the Voronoi diagram [35] based on $N$ reference configurations. These reference configurations are chosen as follows:

1. The first reference configuration is randomly chosen from the current set of $M \times N$ configurations (i.e., $M$ configurations per bin).

2. Suppose we already have $n$ chosen references ($n < N$). For every configuration '$i$', calculate the distances to each of these $n$ previous references, and then find the minimum of these distances, denoted $D_{min}(i)$.

3. The configuration with the maximum $D_{min}(i)$ will be the next reference.

This procedure guarantees that the reference configurations will spread evenly over the represented configurational space. Furthermore the definition of bins will evolve with the simulation as shown in Fig. 5.

We apply this adaptive WE method to a toy two-dimensional system with two distinct pathways. The two-dimensional double well potential is inspired by Chen, Nash and Horing's work, [36] and defined via

$$\begin{aligned} U(x, y)/k_B T &= 20(x^2 + y^2 - 1)^2 y^2 \\ &\quad - \exp\{-4[(x-1)^2 + y^2]\} - \exp\{-4[(x+1)^2 + y^2]\} \\ &\quad + \exp[8(x - 1.5)] + \exp[-8(x + 1.5)] \\ &\quad + 4\exp[-4(y + 0.25)] + 16\exp(-2x^2), \end{aligned} \quad (11)$$

as shown in the left panel of Fig. 4. The barrier between two wells is about $15k_B T$ along each pathway. Both the WE methods with dynamic Voronoi bins and static bins are applied to get the distribution of duration of transition events. The initial state and the final state are defined as the regions where the potential satisfies $U(x, y) < (U_{min} + 2k_B T)$, where $U_{min}$ is the lowest potential in the left and right wells.

We compare the results obtained via WE simulations with dynamic Voronoi bins and static bins in the right graph of Fig. 4. Again, The WE method with dynamic bins reproduces the result from static bins. In Fig. 5, we show the evolution of Voronoi diagram in one of the WE simulations. The bins follow the evolving probability distribution and find the target spontaneously after $\sim 200\tau$, where $\tau$ is the time interval between resamplings (splitting and combination).

## IV. DISCUSSION

### A. How resampling can improve efficiency in WE simulation

A simple example illustrates how WE simulation can improve the efficiency of estimating dynamical quantities like the transition rate or of sampling the transition path ensemble. Consider a simple double-well potential, and imagine starting 100 brute force simulations from the minimum of the left well, state A. At some time $t$ later, one expects to sample approximately $100kt$ transitions to the right well (state B), where $k$ is the rate for that process. For $t \ll 1/k$, it is unlikely to observe any transitions.

Consider also a WE simulation of the same system, using 100 simulations also started from state A, with just a single resampling process at time $t/2$. At such an intermediate time, if the rate is low, we expect that roughly half the trajectories ($\sim 50$) have diffused to the left of the state A minimum and half to the right. Assume the WE resampling process at $t/2$ removes about half of the left trajectories and replicates half of the right trajectories (with appropriate reweighting), maintaining 100 trajectories overall. Because successful transitions must proceed from the right side of the minimum, and because there will now be about $\sim 75$ trajectories on the right instead of $\sim 50$ without resampling, the resampling will increase the chance of observing a transition trajectory. With repeated resampling and a series of bins covering the reaction coordinate, a kind of "statistical ratcheting" is achieved.

There is a price paid for the increased efficiency of estimating dynamical quantities, namely, a *decreased precision* in the sampling of the initial state, A. This is especially true when there is a low transition rate out of state A. However the typical goal of WE simulation is to sample transitions and not the initial state.

### B. Improved efficiency is not guaranteed

On a practical level, we note that WE simulation (i.e., resampling a set of dynamical trajectories) *does not necessarily increase the efficiency with which a certain dynamical process is characterized.* To see this, consider again the example of a double-well potential with trajectories initiated from the left well. If, perversely, the trajectories were resampled so as to have fewer in the transition region between the two wells (and perhaps more to the left of the left minimum), such a WE simulation would likely be less efficient than simple brute force simulation with no resampling. In fact, we found that WE method could be more or less efficient than brute force depending on the particular random walk problem selected (see, Sec. III B).

There is therefore, as with many other simulation strategies, a certain amount of "art" to optimizing the efficiency of the WE method - even though it will remain rigorous if performed based on rigorous resampling. This aspect has been discussed in [3] but bears further investigation.

### C. Connection to other methods

In a sense, we have shown that the WE approach accounts properly for the trajectory "history" and therefore is applicable for non-Markovian dynamics. Other path sampling methods can also account for trajectory history, including "forward flux sampling" (FFS),[4] "dynamic importance sampling" (DIMS),[13] "transition path sampling" (TPS),[12] "transition interface sampling" (TIS),[17] etc.. The Monte Carlo method "Russian roulette and splitting",[8] which is closely related to WE method, apparently also can account for history effects. Failure to account for trajectory history rules out an exact description of non-Markovian processes.

## V. CONCLUSIONS

We used probabilistic arguments and numerical tests to demonstrate the generality of weighted ensemble path sampling simulation. The method, in fact, applies to a broad class of Markovian and non-Markovian dynamics. We also confirmed that the bins used for "sampling" trajectories can be changed during a simulation. In a toy system, we demonstrated an adaptive approach to changing bins which did not require knowledge of the target state.

## VI. ACKNOWLEDGMENTS

We benefited from many helpful discussions about WE approach with group members Divesh Bhatt and Andrew Petersen. This work was supported by NIH Grant GM070987 (to D.M.Z.).

[1] G. A. Huber and S. Kim, Biophys. J. **70**, 97 (1996).

[2] A. Rojnuckarin, S. Kim, and S. Subramaniam, Proc. Natl. Acad. Sci. USA **95**, 4288 (1998).

[3] B. W. Zhang, D. Jasnow, and D. M. Zuckerman, Proc. Natl. Acad. Sci. USA **104**, 18043 (2007).

[4] R. J. Allen, P. B. Warren, and P. R. ten Wolde, Phys. Rev. Lett. **94**, 018104 (2005).

[5] F. A. Escobedo, E. E. Borrero, and J. C. Araque, J. Phys.: Condens. Matter **21**, 333101 (2009).

[6] A. Warmflash, P. Bhimalapuram, and A. R. Dinner, J. Chem. Phys. **127**, 154112 (2007).

[7] A. Dickson, A. Warmflash, and A. R. Dinner, J. Chem. Phys. **130**, 074104 (2009).

[8] H. Kahn, *"Use Of Different Monte Carlo Methods,"* Symposium on Monte Carlo Methods (New York: Wiley, 1956), pp. 146–190.

[9] L. R. Pratt, J. Chem. Phys. **85**, 5045 (1986).

[10] J. Schlitter, M. Engels, P. Krüger, E. Jacoby, and A. Wollmer, Mol. Simul. **10**, 291 (1993).

[11] S. Izrailev, S. Stepaniants, M. Balsera, Y. Oono, and K. Schulten, Biophys. J. **72**, 1568 (1997).

[12] C. Dellago, P. G. Bolhuis, F. S. Csajka, and D. Chandler, J. Chem. Phys. **108**, 1964 (1998).

[13] D. M. Zuckerman and T. B. Woolf, J. Chem. Phys. **111**, 9475 (1999).

[14] P. Eastman, N. Gronbech-Jensen, and S. Doniach, J. Chem. Phys. **114**, 3823 (2001).

[15] W. E, W. Ren, and E. Vanden-Eijnden, Phys. Rev. B **66**, 052301 (2002).

[16] D. J. Wales, Mol. Phys. **100**, 3285 (2002).

[17] T. S. van Erp, D. Moroni, and P. G. Bolhuis, J. Chem. Phys. **118**, 7762 (2003).

[18] A. K. Faradjian and R. Elber, J. Chem. Phys. **120**, 10880 (2004).

[19] Z. Yang, P. MÃjek, and I. Bahar, PLoS Comput. Biol. **5**, e1000360 (2009).

[20] N. Madras and G. Slade, *The Self-Avoiding Walk* (Birkhäuser Boston, 1996).

[21] D. Forster, *Hydrodynamic Fluctuations, Broken Symmetry, And Correlation Functions* (Westview Press, 1995).

[22] D. Bhatt, B. W. Zhang, D. M. Zuckerman, arXiv:0910.5255v1 [physics.bio-ph]

[23] J. S. Liu, *Monte Carlo Strategies in Scientific Computing* (Springer, 2002).

[24] D. Frenkel and B. Smit, *Understanding Molecular Simulation: From Algorithms to Applications*, 2nd ed. (Academic Pr, 2001).

[25] N. G. van Kampen, *Stochastic Processes in Physics and Chemistry* (North Holland, 1992).

[26] M. I. Dykman, P. V. E. McClintock, V. N. Smelyanski, N. D. Stein, and N. G. Stocks, Phys. Rev.

Lett. **68**, 2718 (1992).

[27] R. F. Fox, I. R. Gatland, R. Roy, and G. Vemuri, Phys. Rev. A **38**, 5938 (1988).

[28] R. Crehuet, M. J. Field, and E. Pellegrini, Phys. Rev. E **69**, 012101 (2004).

[29] B. W. Zhang, D. Jasnow, and D. M. Zuckerman, J. Chem. Phys. **126**, 074504 (2007).

[30] F. T. Wall and J. J. Erpenbeck, J. Chem. Phys. **30**, 634 (1959).

[31] P. Grassberger, Phys. Rev. E **56**, 3682 (1997).

[32] P. Grassberger, Comput. Phys. Commun. **147**, 64 (2002).

[33] E. E. Borrero and F. A. Escobedo, J. Chem. Phys. **127**, 164101 (2007).

[34] E. E. Borrero and F. A. Escobedo, J. Chem. Phys. **129**, 024115 (2008).

[35] F. Aurenhammer, ACM Comput. Surv. **23**, 345 (1991).

[36] L. Y. Chen, P. L. Nash, and N. J. M. Horing, J. Chem. Phys. **123**, 094104 (2005).

Figure Captions

Figure 1: The event-duration distribution $\rho_b$ for double well potential $U(x) = 5.0k_BT[1 - x^2]^2$. The left graph shows the results from brute force and WE simulation for different colored noises with $s = 1$, $s = 20$, $s = 50$ and $s = 100$. The right graph shows the WE simulation results for colored noise with $s = 1$ and white noise. For any $s \gg 1$, the white and colored noise results are dramatically different, and WE method reproduces the effects in detail.

Figure 2: "Myopic self-avoiding random walk" on a two-dimensional grid. The left graph shows two successful transition events. The walkers, which started from the origin $(0,0)$, reached the right absorbing wall at $x = 15$ before being absorbed by the left absorbing wall at $x = -1$. The upper trajectory crossed itself once at $(4,10)$, and the lower one avoided itself successfully during the whole transition. The right graph compares the distributions of the transition-event durations, obtained from brute force and WE simulation.

Figure 3: The modified WE program with fluctuating bins $N$ (randomly uniformly chosen from $50 \leq N \leq 100$) is applied to study the duration of transition events for a Brownian particle in double well potential $U(x) = 5.0k_BT[1 - x^2]^2$. The left panel shows the time series of the number of bins $N$; only the first 100 values of $N$ are included. The right panel shows that the modified WE program with fluctuating bins reproduces the results from static bins.

Figure 4: The left graph shows the two-dimensional double well potential defined via Eq.(11) and the potential values of $U/k_BT$ at extreme points. The right graph compares the results of duration of transition event duration obtained by the WE methods with dynamic Voronoi bins and static bins. They are in good agreement.

Figure 5: An adaptive WE method which defines Voronoi bins based on the evolving distribution The dots are the reference configurations for the Voronoi bins. At the beginning, all the reference configurations are in the initial state. Subsequently the reference configurations and bins follow the evolving probability distribution. The target is found spontaneously after $200\tau$, where $\tau$ is the time interval between resamplings (splitting and combination).

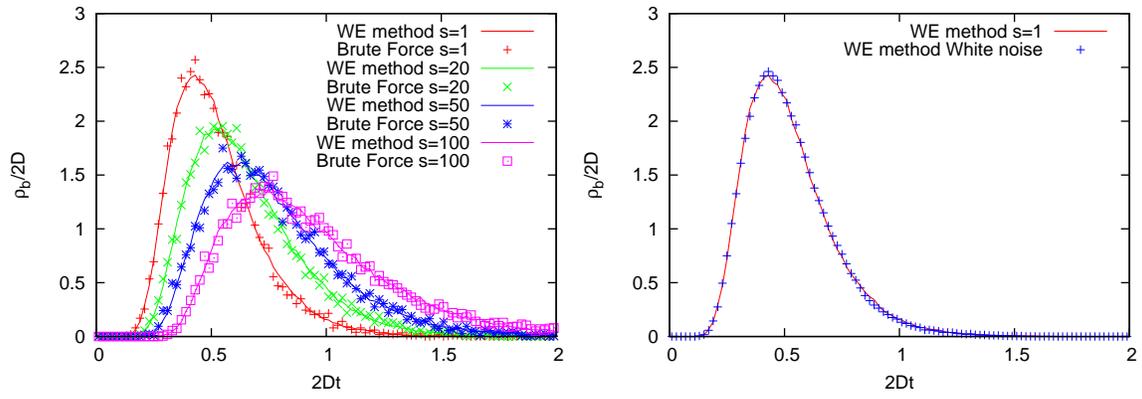

Figure 1:

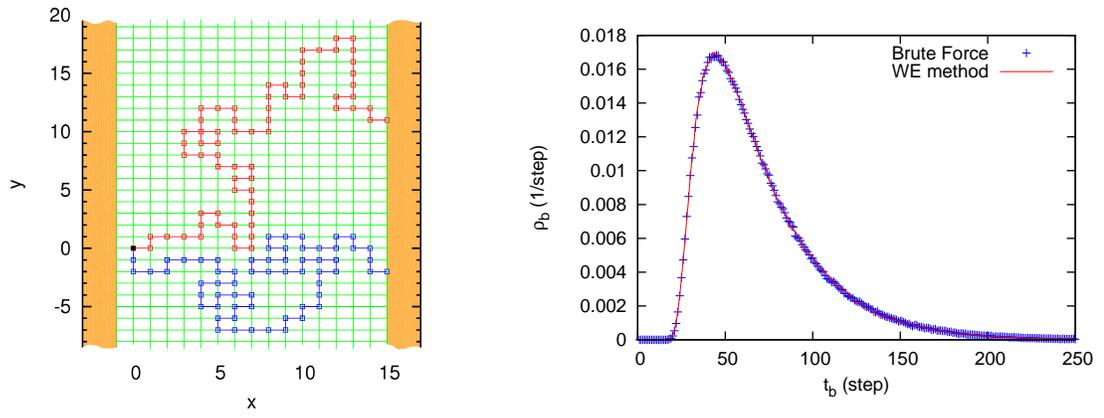

Figure 2:

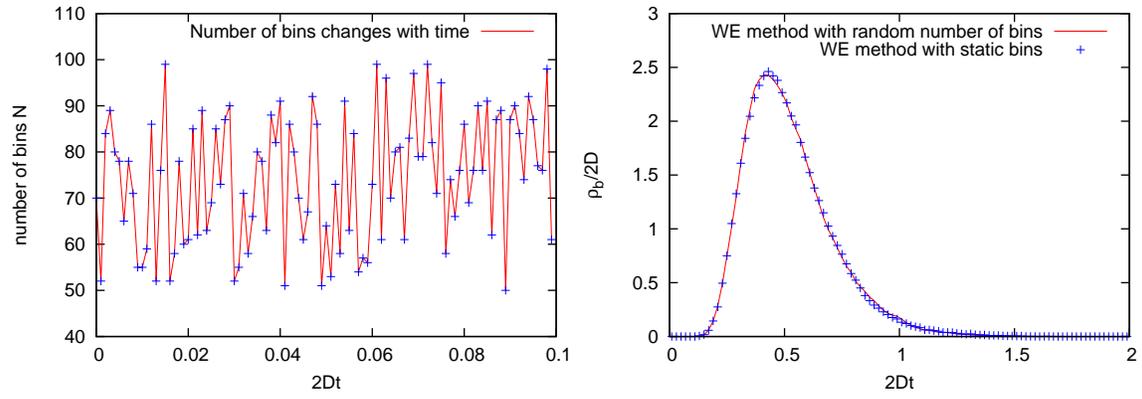

Figure 3:

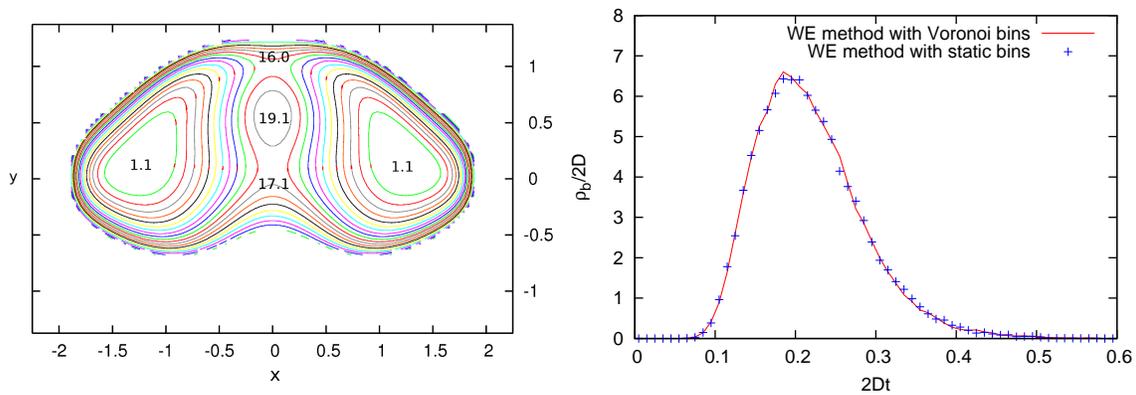

Figure 4:

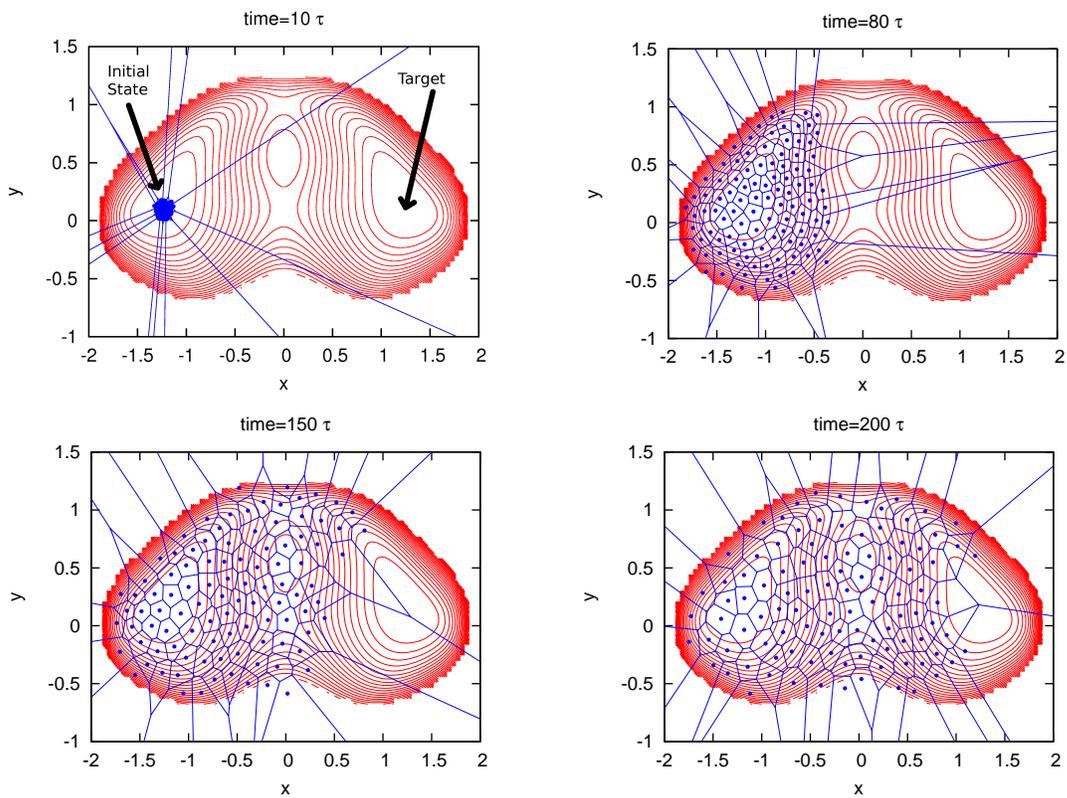

Figure 5: